\documentclass[sigconf,dvipsname,authorversion]{acmart}

\usepackage{datetime}
\usepackage{tikz}
\usepackage{pdfpages}
\usepackage{amsmath}

\newtheorem{property}{Property}

\usepackage{enumitem}
\usepackage[font=small,labelfont=bf]{caption}
\usepackage[font=small]{subcaption}
\usepackage{graphicx}
\setlistdepth{15}
\newlist{caselist}{enumerate}{10}
\setlist[caselist]{label=\roman*.}
\usetikzlibrary{positioning,shapes}
\tikzstyle{block} = [draw, thick, fill=gray!20, rounded corners, minimum height=2em, minimum width=6em]

\copyrightyear{2021}
\acmYear{2021}
\setcopyright{acmlicensed}\acmConference[ICDCN '21]{International Conference on Distributed Computing and Networking 2021}{January 5--8, 2021}{Nara, Japan}
\acmBooktitle{International Conference on Distributed Computing and Networking 2021 (ICDCN '21), January 5--8, 2021, Nara, Japan}
\acmPrice{15.00}
\acmDOI{10.1145/3427796.3427815}
\acmISBN{978-1-4503-8933-4/21/01}

\begin{document}

\title{Fast Flexible Paxos: Relaxing Quorum Intersection for Fast Paxos}

\author{Heidi Howard}
\affiliation{%
  \institution{University of Cambridge}
}
\email{heidi.howard@cl.cam.ac.uk}

\author{Aleksey Charapko}
\affiliation{%
  \institution{University of New Hampshire}
}
\email{aleksey.charapko@unh.edu}

\author{Richard Mortier}
\affiliation{%
  \institution{University of Cambridge}
}
\email{richard.mortier@cl.cam.ac.uk}

\begin{abstract}
Paxos, the de facto standard approach to solving distributed consensus, operates in two phases, each of which requires an intersecting quorum of nodes.
Multi-Paxos reduces this to one phase by electing a leader but this leader is also a performance bottleneck.
Fast Paxos bypasses the leader but has stronger quorum intersection requirements.

In this paper we
observe that Fast Paxos' intersection requirements can be safely relaxed, reducing to just one additional intersection requirement between phase-1 quorums and any pair of fast round phase-2 quorums.
We thus find that the quorums used with Fast Paxos are larger than necessary, allowing alternative quorum systems to obtain new tradeoffs between performance and fault-tolerance.
\end{abstract}


\maketitle

\section{Introduction}
\label{sec:intro}



Paxos~\cite{lamport_tcs98,lamport_sigact01} and its variants~\cite{lamport_msr05,camargos_podc07,yanhua_osdi08,sutra_srds11, moraru_sosp13,ongaro_atc14} provide reliable solutions to the problem of distributed consensus~\cite{fischer_jacm85}.
Thanks to their excellent fault-tolerance properties and proven consistency guarantees, these algorithms often underpin the replicated state machines~\cite{schneider_cs90} at the heart of many industrial distributed systems, e.g.,~Chubby~\cite{burrows_osdi06}, CockroachDB~\cite{taft_sigmod20}, and PaxosStore~\cite{zheng_vldb17}.

Traditionally, the Paxos family of algorithms uses majority quorums, guaranteeing that any two sets containing the majority of nodes intersect, ensuring that previously decided values are not lost.
Flexible Paxos~\cite{howard_opodis16} relaxes the requirement for intersecting quorums in Paxos, proving that quorum intersection is only required between phases, permitting disjoint quorums to be used within each phase.
This result enabled subsequent algorithms to improve performance by adjusting quorums depending on the phase of the algorithm~\cite{ailijiang_tpds20, nawab_icmd18,muhammed_nsdi19,enes_eurosys20,eischer_popoc20,logdevice}.

Paxos is usually implemented using Multi-Paxos~\cite{lamport_tcs98,lamport_sigact01}, an optimization that elects one node to be a \emph{leader}.
This single leader can then achieve distributed consensus in just one phase, but unfortunately also becomes a performance bottleneck.

Seeking to improve performance, a new family of \emph{leaderless} consensus algorithms emerged, starting with Fast Paxos~\cite{lamport_msr05}, which forms the basis for subsequent algorithms including Generalized Paxos~\cite{lamport_msr04} and Egalitarian Paxos~\cite{moraru_sosp13}.
Paxos uses the idea of \emph{rounds} in which at most one value can be \emph{proposed}.
Fast Paxos introduced the notion of \emph{fast rounds} where multiple values can be safely proposed in the same round.
However, such fast rounds require stronger quorum intersection than classical rounds.
Specifically, Paxos only requires that any two quorums intersect whereas Fast Paxos also requires that any quorum intersects with any two fast round quorums.
Fast Paxos' quorum intersection requirements can be satisfied by requiring fast round quorums to contain at least three-quarters of nodes.
Due to this additional quorum intersection requirement, Fast Paxos and its variants cannot directly benefit from Flexible Paxos.

In this paper, we show that the approach of Flexible Paxos can be safely applied to consensus algorithms that rely on stronger quorum intersection by extending Flexible Paxos to Fast Paxos.
The resulting algorithm, which we refer to as \emph{Fast Flexible Paxos}, relaxes the quorum intersection requirements of Fast Paxos.
Specifically, Fast Flexible Paxos proves that the only additional quorum intersection requirement is between phase-1 quorums and any pair of fast round phase-2 quorums.



Relaxed quorum intersection in Fast Flexible Paxos permits new performance tradeoffs by manipulation of the quorum systems.
For example, reducing the size of fast round quorums reduces the contention in the algorithm.
This may further improve overall performance on top of an improvement attained by a smaller quorum alone.
For instance, we illustrate that the Fast Flexible Paxos with smaller fast quorums achieves up to 10\% better latency than Fast Paxos in low conflict scenarios.

\section{Background}
\label{sec:background}

\begin{figure}
  \centering
\begin{tikzpicture}[node distance= 0.8cm and 2.8 cm, on grid]
    \node [block] (paxos) {Paxos};
    \node [block, above right = of paxos] (flex) {Flexible Paxos};
    \node [block, above left = of paxos] (fast) {Fast Paxos};
    \node [block, above left = of flex, above right = of fast] (fastflex) {Fast Flexible Paxos};
    \draw [<-,thick] (paxos) -- (flex);
    \draw [<-,thick] (paxos) -- (fast);
    \draw [<-,thick] (paxos) -- (fastflex);
    \draw [<-,thick] (fast) -- (fastflex);
    \draw [<-,thick] (flex) -- (fastflex);
\end{tikzpicture}
\caption{Relationships between distributed consensus algorithms. The arrows denote where one algorithm generalizes over another.}
\label{fig:algorithms}
\end{figure}
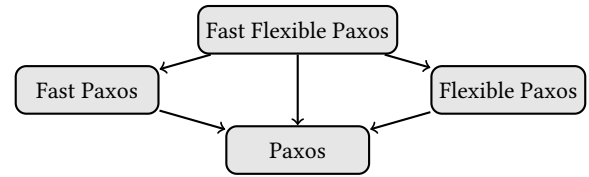

We begin by recapping how distributed consensus is currently solved by Paxos, Flexible Paxos, and Fast Paxos.
The relation between these algorithms is shown in Figure~\ref{fig:algorithms}.

\subsection{Paxos}
The Paxos algorithm distinguishes between two roles a node can take: a \emph{proposer} and an \emph{acceptor}.
A proposer initiates a decision by executing Paxos using a \textit{round}.
Rounds are integers allocated to proposers, and each proposer must propose only one value in each round. A proposal is a pair of a round and a value.
The algorithm runs in two phases per round, each requiring a quorum of acceptors to proceed.

In phase-1, the proposer learns if a value was decided in any previous round by asking the acceptors to send the last proposal they voted for.
Acceptors promise not to vote in any smaller round.

In phase-2, the proposer asks acceptors to vote for a value $v$.
If during phase-1 the proposer learned that a value might become decided, then it must use that value for $v$.
Provided that an acceptor has not promised otherwise, it updates its last proposal voted for and acknowledges that to the proposer.
Once the proposer has completed phase-2, it will learn that $v$ is decided.

Paxos requires intersection between any two quorums.
If $\mathcal{Q}$ denotes the set of quorums then this intersection requirement can be expressed as:
\begin{equation}
    \forall Q, Q' \in \mathcal{Q}: Q \cap Q' \neq \{\}
\label{eq:paxos}
\end{equation}
Consider a simple quorum system based solely on the cardinality of quorums.
If $q$ denotes the cardinality of quorums in $\mathcal{Q}$ and $n$ denotes the number of acceptors, then we can express Paxos' quorum intersection requirement (Eq.~\ref{eq:paxos}) as:
\begin{equation}
    2q > n
\label{eq:paxos_simple}
\end{equation}

Paxos is often used to decide a sequence of values, where the $i^{\footnotesize\hbox{th}}$ instance of Paxos decides the $i^{\footnotesize\hbox{th}}$ value.
\emph{Multi-Paxos} improves the algorithm's performance by pre-executing the phase-1 of every instance by the same stable proposer, known as a \emph{leader}.
The leader can then decide each value in just two communication steps, compared to the four communication steps needed by Paxos.

\subsection{Flexible Paxos}
Flexible Paxos differentiates between the quorums for phase-1, $\mathcal{Q}_1$ and phase-2, $\mathcal{Q}_2$.
This approach allows the weakening of Paxos' quorum intersection requirement (Eq.~\ref{eq:paxos}) to the following:
\begin{equation}
    \forall Q \in \mathcal{Q}_1, \forall Q' \in \mathcal{Q}_2: Q \cap Q' \neq \{\}
\label{eq:flexpaxos}
\end{equation}
That is, quorum intersection is not required within each of the two phases.
As previously, if $q_{i}$ denotes the cardinality of quorums in $\mathcal{Q}_{i}$, then we can express Flexible Paxos' quorum intersection requirement (Eq.~\ref{eq:flexpaxos}) as:
\begin{equation}
    q_1 + q_2 > n
\label{eq:flexpaxos_simple}
\end{equation}
In Multi-Paxos, phase-1 is executed rarely compared to phase-2 so applications can decrease the phase-2 quorum to improve performance at the cost of decreased fault-tolerance.

\subsection{Fast Paxos}
In Multi-Paxos, the leader is a bottleneck and other proposers must first send values to the leader, adding a communication step.
Fast Paxos addresses these issues by allowing all proposers to propose values directly to the acceptors.
Fast Paxos can thus decide a value in one phase, an optimal solution to distributed consensus~\cite{lamport_tr03}.

Fast Paxos achieves this by introducing \textit{fast} rounds, where it is safe for multiple values to be proposed in the same round.
In a fast round, if the leader is free to propose any value in phase-2 then instead of proposing a specific value it proposes a special \textit{any} value to the acceptors.
Proposers can then send proposals directly to the acceptors and each acceptor will vote for the first proposal it receives as if it had been sent by the leader.

Fast Paxos needs stronger quorum intersection for fast rounds to ensure safety and progress in the case of conflicts.
If $\mathcal{Q}_f$ denotes fast round quorums and $\mathcal{Q}_c$ denotes classic round quorums then Fast Paxos requires:\footnote{The original paper asserts that (a)~any two quorums intersect and (b)~any two fast quorums and any classic or fast quorum intersect. Since (b) already covers the intersection between two fast quorums as well as a fast and classic quorum, we have reduced (a) to any two classic quorums intersect.}
\begin{equation}
    \forall Q,Q' \in \mathcal{Q}_c: Q \cap Q' \neq \{\}
\label{eq:fastpaxos_classic}
\end{equation}
\begin{equation}
    \forall Q,Q' \in \mathcal{Q}_f, \forall Q'' \in \mathcal{Q}_c: Q \cap Q' \cap Q'' \neq \{\}
\label{eq:fastpaxos_classicfast}
\end{equation}
\begin{equation}
    \forall Q,Q',Q'' \in \mathcal{Q}_f: Q \cap Q' \cap Q'' \neq \{\}
\label{eq:fastpaxos_fast}
\end{equation}

In other words, Fast Paxos requires that any pair of classic round quorums intersect (Eq.~\ref{eq:fastpaxos_classic}), that any pair of fast round quorums intersect with any classic round quorum (Eq.~\ref{eq:fastpaxos_classicfast}) and that any three fast round quorums intersect (Eq.~\ref{eq:fastpaxos_fast}).

As before, let $q_{f}$ and $q_{c}$ denote the cardinality of quorums in $\mathcal{Q}_{f}$ and $\mathcal{Q}_{c}$ respectively. We can now express Fast Paxos' quorum intersection requirements (Eq.~\ref{eq:fastpaxos_classic}, Eq.~\ref{eq:fastpaxos_classicfast} \& Eq.~\ref{eq:fastpaxos_fast}) as:
\begin{equation}
2q_c > n
\label{eq:fastpaxos_classic_simple}
\end{equation}
\begin{equation}
q_c + 2q_f > 2n
\label{eq:fastpaxos_classicfast_simple}
\end{equation}
\begin{equation}
3q_f > 2n
\label{eq:fastpaxos_fast_simple}
\end{equation}
Fast Paxos suggests using $q_c=q_f=\lfloor \frac{2n}{3}\rfloor +1$ or $q_c=\lfloor \frac{n}{2}\rfloor +1$ and $q_f=\lceil \frac{3n}{4}\rceil$ to satisfy these requirements.\footnote{The former is sometimes written as $n=3f+1$ and $q_f=q_c=2f+1$ where $f$ is the number of faults which can be tolerated~\cite{junqueira_hotdep07}.}
The larger quorums required by Fast Paxos have been shown to significantly decrease performance compared to Paxos~\cite{junqueira_hotdep07}.

\section{Fast Flexible Paxos}
\label{sec:main}

Following the approach of Flexible Paxos, we will differentiate between the quorums used for each phase of Fast Paxos.

Recall that quorum intersection is required between the two phases to ensure that a proposer learns in phase-1 any value which may be decided in phase-2.
In phase-2, the proposer needs to pick a value from the highest round it learned during phase-1.
In Fast Paxos, there may be multiple such values as acceptors may vote for different values during the phase-2 of fast rounds, requiring a proposer to determine which single value (if any) could be decided in previous rounds.
Fast Flexible Paxos achieves this by ensuring that each phase-1 quorum intersects with any pair of fast round phase-2 quorums.

The quorum intersection requirements are the same regardless of whether phase-1 is for a fast or classic round.
Therefore $\mathcal{Q}_1$ denotes the phase-1 quorums (fast or classic) whereas $\mathcal{Q}_{2c}$ and $\mathcal{Q}_{2f}$ denote the phase-2 quorums for classic and fast rounds respectively.
The weakened intersection requirements for Fast Paxos are as follows:
\begin{equation}
\forall Q \in \mathcal{Q}_1, \forall Q' \in \mathcal{Q}_{2c}: Q \cap Q' \neq \{\}
\label{eq:fastflexpaxos_classic}
\end{equation}
\begin{equation}
\forall Q \in \mathcal{Q}_1, \forall Q',Q'' \in \mathcal{Q}_{2f}: Q \cap Q' \cap Q'' \neq \{\}
\label{eq:fastflexpaxos_fast}
\end{equation}

In other words, we find that quorum intersection is only required between phase-1 quorums and phase-2 classic round quorums (Eq.~\ref{eq:fastflexpaxos_classic}) and between phase-1 quorums and any pair of phase-2 fast round quorums (Eq.~\ref{eq:fastflexpaxos_fast}).
Note that quorum intersection is not required between phase-1 quorums, between phase-2 classic round quorums, or  between phase-2 classic round quorums and phase-2 fast round quorums.

If $q_1$, $q_{2f}$ and $q_{2c}$ denote the cardinality of quorums in $\mathcal{Q}_1$, $\mathcal{Q}_{2f}$ and $\mathcal{Q}_{2c}$ respectively then we can express Fast Flexible Paxos' intersection requirements (Eq.~\ref{eq:fastflexpaxos_classic} \& Eq.~\ref{eq:fastflexpaxos_fast}) as:
\begin{equation}
q_1 + q_{2c} > n
\end{equation}
\begin{equation}
q_1 + 2q_{2f} > 2n
\end{equation}

\section{Correctness}

Fast Flexible Paxos must ensure that at most one value is decided.
We can show this by proving the following two properties:

\begin{property}
At most one value is decided per round.
\label{prop:safe_within_rounds}
\end{property}

\begin{proof}
If the round is classic, then at most one proposer can propose (and therefore decide) a value.
If the round is fast, then at most one value will be decided as any two fast phase-2 quorums will intersect (Eq.~\ref{eq:fastflexpaxos_fast}).
\end{proof}

\begin{property}
A proposer will only propose a value in a given round if no smaller round can decide a different value.
\label{prop:safe_across_rounds}
\end{property}

\begin{proof}
Assume that a value $v$ is decided in round $r$.
Consider the next round $r'$ ($r'>r$) where a value is proposed.
In phase-1, the proposer of round $r'$ will ask the acceptors to promise not to vote in any smaller rounds and to reply with the last proposal they voted for.

Due to quorum intersection between the phase-1 and phase-2 (Eq.~\ref{eq:fastflexpaxos_classic} \& Eq.~\ref{eq:fastflexpaxos_fast}), at least one acceptor will reply to the proposer in round $r'$ with value $v$ and round $r$.
This is because the acceptor must have voted for value $v$ in round $r$ before participating in round $r'$ as it promises not to vote in any smaller rounds.
The acceptor also cannot have voted in any round since $r$ as round $r'$ is the first round after $r$ where a value is proposed.
The proposer in round $r'$ will propose the value $v$ as it will not receive any proposals from rounds greater than $r$.
If the round $r$ is fast then the proposer in round $r'$ may receive multiple values from round $r$, however, the proposer will choose value $v$ since $v$ is decided.

Consider the next round $r''$ ($r''>r'$) where a value is proposed.
The value proposed in round $r''$ must be value $v$ as only the value $v$ has been proposed since value $v$ was decided in round $r$.
By induction, we can see that for all rounds larger than $r$, if a value is proposed then that value will be $v$.
\end{proof}

Fast Flexible Paxos must also ensure liveness to solve distributed consensus, and in particular, it must satisfy the following property:

\begin{property}
Upon completion of phase-1, a proposer can determine at least one value which is safe to propose in phase-2.
\end{property}

\begin{proof}
Consider a proposer that has just completed phase-1 of round $r'$.
A value is safe to propose in round $r'$ only if the proposer knows that no smaller round has decided a different value (Property~\ref{prop:safe_across_rounds}).
If a proposer receives multiple proposals in phase-1 then it proposes the value with the greatest round $r$ ($r<r'$).
However, if the round $r$ is a fast round the proposer may receive multiple values and so must determine which of the values (if any) could be decided in round $r$.
Note that a value could be decided in round $r$ only if there exists a phase-2 fast round quorum of acceptors which may have voted for the value in round $r$.

For every pair of phase-2 fast round quorums, at least one acceptor which will reply to a proposer in phase-1 of round $r'$ must also vote in both quorums if both quorums decide a value in round $r$ (Eq.~\ref{eq:fastflexpaxos_fast}).
The acceptor will only vote for one value in round $r$ and thus will reply to the proposer with only one value.
The proposer thus learns that the other value cannot have been decided in round $r$ by any quorum containing that acceptor.

Once the proposer has heard from a phase-1 quorum of acceptors, the proposer can safely eliminate either all or all but one of the values received with round $r$.
\end{proof}

In Appendix~\ref{sec:spec}, we adapted the Fast Paxos specification~\cite{lamport_msr05} to model check a formal specification of Fast Flexible Paxos using TLA+~\cite{lamport_tla}.
Both our specification and model checking configurations are also available online~\cite{ffpaxos_tla}.

\section{Implications}
\label{sec:impl}

The weakened intersection requirements show that phase-1 of a fast round can use the same quorum as phase-1 of a classic round.
Since the requirement of fast round quorums is stricter than classic round quorums then fast round quorums must be at least as large as classic round quorums.

For example, Fast Paxos suggests using $q_f=\lceil \frac{3n}{4}\rceil$ and $q_c=\lfloor \frac{n}{2}\rfloor +1$, but our relaxed intersection requirements demonstrate that a simple majority of acceptors is sufficient for phase-1 of fast rounds.
Similarly, Fast Paxos also suggests using $q_c=q_f=\lfloor \frac{2n}{3}\rfloor +1$ and again we observe that is conservative and only one third of acceptors are needed for phase-2 of classic rounds ($q_{2c}=\lceil \frac{n}{3}\rceil$).

More generally, by weakening the intersection requirements of Fast Paxos, we provide more flexibility to choose quorum systems and tradeoffs.
In a stable system, phase-1 is rarely executed compared to phase-2 so we can decrease the size of our phase-2 quorums, fast and classic, provided we increase the size of our phase-1 quorums.
For example, a system of 11 acceptors could use phase-2 quorums of 7 acceptors for fast rounds and 3 acceptors for classic rounds, if it uses quorums of 9 acceptors for phase-1.

Note that the liveness of such a system does depend upon both phase-1 and phase-2 quorums.
For example, we could minimize fast round phase-2 by using a simple majority for $q_{2f}$, but this would require all acceptors to start a new round.

\section{Preliminary Evaluation}

\begin{figure}
    \centering
    \begin{subfigure}{0.47\textwidth}
        \includegraphics[width=\linewidth]{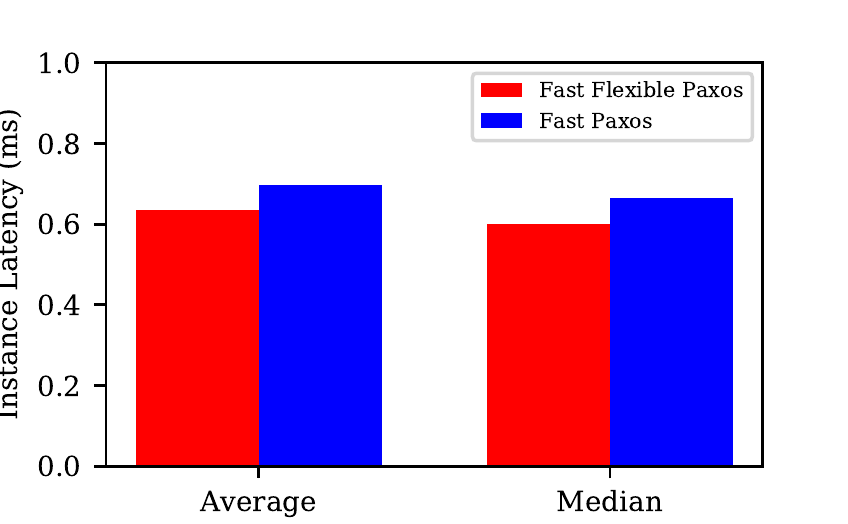}
        \caption{Instance latency in conflict-free workload at 1400 requests/s.}
        \label{fig:paxi_no_conflict}
    \end{subfigure}
    \hspace*{\fill} 
    \begin{subfigure}{0.47\textwidth}
        \includegraphics[width=\linewidth]{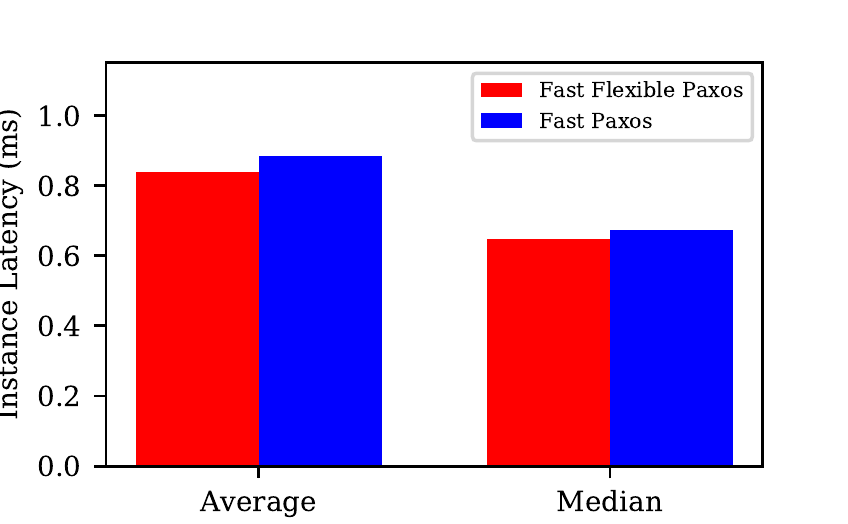}
        \caption{Instance latency in a workload with 0.5\%-1.5\% conflicts at 2700 requests/s.}
        \label{fig:paxi_conflict_latency}
    \end{subfigure}
    \hspace*{\fill} 
    \begin{subfigure}{0.47\textwidth}
        \includegraphics[width=\linewidth]{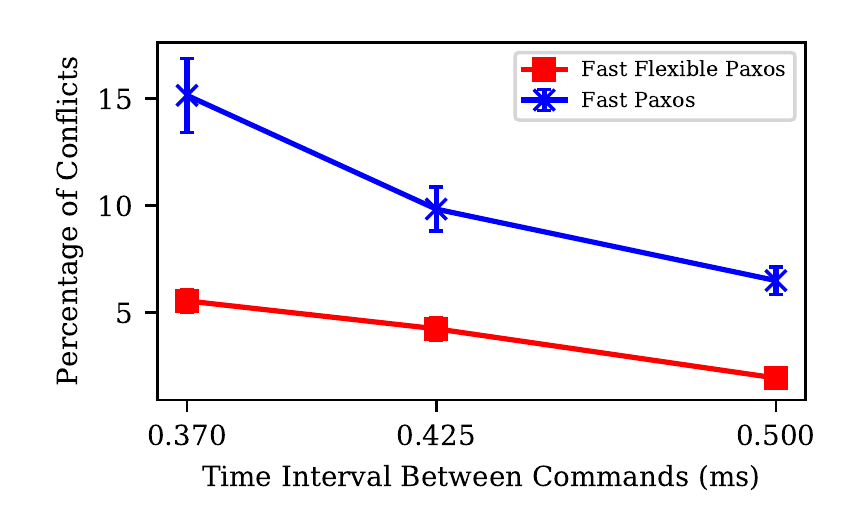}
        \caption{Probability of conflict under different intervals between potentially conflicting commands.}
        \label{fig:paxi_conflict}
    \end{subfigure}

    \caption{Evaluation of Fast Flexible Paxos in a Paxi 11 node cluster. Fast Paxos uses quorums of the following sizes: $q_c=6$ and $q_f=9$.  Fast Flexible Paxos runs with $q_1=9$, $q_{2f}=7$, and $q_{2c}=3$ quorums. }
\label{fig:paxi}
\end{figure}

The main contribution of this paper is the observation that the quorum intersection requirements of Fast Paxos can safely be relaxed.
We have also implemented Fast Flexible Paxos to illustrate the potential performance improvements this result enables, even with a simple quorum system based solely on quorum cardinality.

We evaluated Fast Flexible Paxos with the aforementioned quorum configuration ($q_1=9$, $q_{2f}=7$, and $q_{2c}=3$) using Paxi~\cite{ailijiang_icmd19} on AWS EC2 m5a.large VMs. We focused on two key aspects of Fast Flexible Paxos: latency and conflict reduction due to the smaller phase-2 fast round quorums.
We compared our algorithm against a Fast Paxos ($q_c = 6$ and $q_f = 9$) baseline.

In Figure~\ref{fig:paxi_no_conflict} we illustrate the performance of two algorithms under a workload of 1400 requests/second with no conflicts.
Smaller fast round quorums allowed Fast Flexible Paxos to reduce the average and median latency by 5---8\% compared to Fast Paxos.

We also evaluated under conflicts by generating a workload with several clients racing to propose different commands for the same consensus instance.
For this workload we generated a steady stream of operations with only small intervals between them.
We also pre-assigned each operation to an instance to control the potential for conflicts.
In about 10\% of the cases, we assigned the same instances to two consecutive operations, creating a race condition between them.
In this setup there are two possible outcomes for such races: one of the operations reaches the fast round quorum, causing the second to abort; or none of the operations reach the fast round quorum, causing entry into the conflict resolution phase.
We then measured the conflict avoidance ratio to study the impact of the Fast Flexible Paxos on conflict handling.

We found that Fast Flexible Paxos entered the conflict recovery almost one-third as frequently as Fast Paxos due to the smaller fast quorum.
However, the overall frequency of recovery phases increased substantially for both algorithms as the throughput rises and the interval between the commands shrinks, as Figure~\ref{fig:paxi_conflict} shows.
Considering the overall performance in the conflict workload, Figure~\ref{fig:paxi_conflict_latency} shows that our Fast Flexible Paxos continues to maintain a roughly 5\% latency advantage over Fast Paxos even under high load compared to our non-conflict experiment.

We believe Fast Flexible Paxos will enable further performance improvements if  quorum systems are used that are not based  solely on quorum cardinality~\cite{gifford_sosp79,garcia_jacm85,peleg_podc95,naor_siam98,junqueira_hotdep05}.
This has already proven to be the case for Flexible Paxos~\cite{ailijiang_tpds20, nawab_icmd18,muhammed_nsdi19,enes_eurosys20,eischer_popoc20,logdevice}.
In particular, Fast Flexible Paxos can benefit from the existing literature on Byzantine and Refined quorum systems~\cite{malkhi_dc98,guerraoui_podc07} as these quorum systems provide stronger quorum intersection.

\section{Summary}
\label{sec:summary}

Fast Paxos allows any proposer to decide a value in two communication steps in the absence of collisions.
This is the optimal number of communication steps for distributed consensus.
However, to achieve this it needs a stronger quorum intersection than Paxos and thus has not benefited from recent work on relaxing quorum intersection requirements.

Fast Flexible Paxos weakens Fast Paxos' quorum intersection requirements by differentiating between the quorums used in each phase of the algorithm.
We find that quorum intersection is only required between any phase-1 quorum and both (a) any phase-2 classic round quorum, and (b) any pair of phase-2 fast round quorums.
This shows that the quorum systems used by Fast Paxos are conservative and that alternative quorum systems could be safely used.

More generally, we have proven that the approach of Flexible Paxos generalizes to distributed consensus algorithms beyond Paxos.
We hope more consensus algorithms, particularly those which extend Fast Paxos such as Generalized Paxos~\cite{lamport_msr04}, Egalitarian Paxos~\cite{moraru_sosp13}, MDCC~\cite{kraska_eurosys13}, Alvin~\cite{turcu_opodis14} and Caesar~\cite{arun_dsn17}, adopt this approach to relax their quorum intersection requirements, giving applications greater flexibility to determine their performance and fault-tolerance tradeoffs.

\begin{acks}
This work was funded in part by EPSRC EP/M02315X/1.
\end{acks}

\newpage
\bibliographystyle{ACM-Reference-Format}
\bibliography{refs}

\appendix


\section{Formal Specification of Fast Flexible Paxos}
\label{sec:spec}

This appendix presents a formal specification of Fast Flexible Paxos, written in TLA+ and model checked in TLC~\cite{lamport_tla}.
Both the TLA+ specification and TLC configuration are available online~\cite{ffpaxos_tla}.

This specification is the result of only a minor modifications to the original Fast Paxos specification~\cite{lamport_msr05}.
Underlined comments highlight where changes have been made to the original specification, particularly regarding the distinction between phase-1 and phase-2 quorums.
Readers may wish to pay particular attention to $IsPickableVal(Q,i,M,v)$ which implements the proposer's rule for picking a value $v$ to send to acceptors in phase-2 of round $i$ after receiving the messages $M$ from the acceptor quorum $Q$.
This specification refers to proposers as coordinators.

\onecolumn
\newcount\page
\page=1
\loop
\centering
\includegraphics[trim=4cm 4cm 4cm 3cm ,clip,page=\page]{FastFlexiblePaxos.pdf}
\newpage
\advance \page 1
\ifnum \page<11
\repeat

\end{document}